\documentclass[aps,prl,reprint,groupedaddress,longbibliography]{revtex4-2}

\begin{document}


\title{Reply to ``Comment on `Directed motion of spheres by unbiased driving forces in viscous fluids beyond the Stokes' law regime' ''}


\author{Jes\'us Casado-Pascual}
\email[]{jcasado@us.es}
\affiliation{F\'{\i}sica Te\'orica, Universidad de Sevilla, Apartado de Correos 1065, Sevilla 41080, Spain}


\date{\today}

\begin{abstract}
We refute the criticism expressed in a Comment by P. J. Mart\'{\i}nez and R. Chac\'on [arXiv:2105.09308] on our paper Phys. Rev. E \textbf{97}, 032219 (2018). We first show that our paper is free of any mistakes and that the criticism expressed in the Comment is based on  a misunderstanding or misrepresentation of what is actually claimed in our paper. We also show that the argument presented against the general formalism mentioned in the Comment is invalid because it is based on a confusion by the authors of the Comment between the general formalism per se and its lowest order approximation. In fact, the arguments contained in the Comment do not show that such a formalism is in general incorrect, as the authors of the Comment suggest, but rather that the supposedly universal predictions of the theory of ratchet universality are in general incorrect. In addition, we show that other arguments presented in Comment are invalid because they are based on incorrect equations.
\end{abstract}


\maketitle

In our recent paper~\cite{casado:2018}, we investigate the generation of directed motion in a system consisting of a sphere immersed in a viscous fluid and subjected to time-periodic forces of zero average. Specifically: (i) we obtain necessary conditions for the existence of such directed motion, (ii) we derive an analytical expression for the average terminal velocity of the sphere in the adiabatic limit, (iii) we carry out a numerical analysis of the dependence of the average terminal velocity on the system parameters and compare the results with those obtained using the adiabatic approximation, and (iv) we explain some aspects of the observed phenomenology by means of symmetry arguments. It is important to emphasize that none of these results is questioned by Mart\'{\i}nez and Chac\'on (abbreviated hereinafter as M\&C) in their Comment~\cite{martinez:2021}. On the contrary, comparison of the top and bottom panels of Figs.~1 and 2 of~\cite{martinez:2021}, as well as of the top and middle panels of Fig.~3 of~\cite{martinez:2021}, only confirms the validity of the analytical expression for the average terminal velocity in the adiabatic limit derived in the commented paper [Eq.~(12) of~\cite{casado:2018}]. Note that this analytical expression is wrongly written in the Comment [compare Eq.~(3) of~\cite{martinez:2021} with Eq.~(12) of~\cite{casado:2018}]. 

M\&C focus their criticism mainly on the following text that appears below Fig.~3 of~\cite{casado:2018}: ``The curves in Fig.~3 also reveal that, for fixed values of the other parameters, there exists an optimal value of $\zeta$ which maximizes the second component of the average terminal velocity. Furthermore, as $\omega \tau$ increases, the maximum velocity decreases and its location shifts towards lower values of $\zeta$. It should be noted here that, in the lowest order, the general formalism developed in Refs.~[27, 28]'' (for Refs.~\cite{cuesta:2013,casado:2015}) ``leads to the approximate expression $\overline{V}_2(\zeta)\approx C \zeta^2(1-\zeta)$, where $C$ is independent of $\zeta$. This expression vanishes at $\zeta=0$ and $\zeta=1$, and displays a maximum at $\zeta=2/3$, thus qualitatively resembling the behavior seen in Fig.~3. However, it is unable to account for the dependence of the location of the maximum velocity on $\omega \tau$. This deficiency is not surprising, given that the above approximation is expected to be accurate only for small values of $f_0$ and, in Fig.~3, we have taken $f_0=100$.'' In reference to this text, M\&C claim ``\textit{This Comment will question some of the above statements.}'', but they do not explicitly indicate which of the above statements they question.
 
It should be pointed out that the aforementioned text of~\cite{casado:2018} only describes objective facts or easily verifiable mathematical facts.
Indeed, the first two sentences provide an objective description of what is observed in Fig.~3 of~\cite{casado:2018}. The third sentence indicates a mathematical fact that can easily be verified, namely, ``in the lowest order, the general formalism developed in Refs.~[27,~28]'' (for Refs.~\cite{cuesta:2013,casado:2015}) ``leads to the approximate expression $\overline{V}_2(\zeta)\approx C \zeta^2(1-\zeta)$, where $C$ is independent of $\zeta$''. The fourth sentence indicates some properties of the function $C \zeta^2(1-\zeta)$ that can  easily be verified, namely, the function $C \zeta^2(1-\zeta)$  vanishes at $\zeta=0$ and at $\zeta=1$, displays a maximum at $\zeta=2/3$, and its shape qualitatively resembles the behavior seen in Fig.~3 of~\cite{casado:2018}. The fifth sentence indicates an objective fact, namely,  the approximate expression $\overline{V}_2(\zeta)\approx C \zeta^2(1-\zeta)$ ``is unable to account for the dependence of the location of the maximum velocity on $\omega \tau$''.  Finally, the sixth sentence indicates a well-known mathematical fact, namely, in a series expansion (in the present case, the series expansion obtained by applying the general formalism developed in~\cite{cuesta:2013,casado:2015}), the lowest-order approximation is expected to be accurate only for small values of the expansion parameter (in the present case, for small values of $f_0$) and, consequently, it is not surprising that such an approximation is inadequate if the expansion parameter is large (in the present case, $f_0=100$). None of these statements is questioned in the Comment~\cite{martinez:2021}.

Referring again to the aforesaid text of~\cite{casado:2018}, M\&C claim in the Abstract ``\textit{The author explains the dependence on the relative amplitude of the two harmonic components of the average terminal velocity from the perspective of a general formalism. In this Comment, this explanation is shown to be in general incorrect, \dots}'' and at the end of the Comment ``\textit{In conclusion, the theoretical explanation discussed in Ref.~[1] is in  general incorrect, \dots}''. It is important to emphasize that at no point in~\cite{casado:2018} do we claim to explain ``\textit{the dependence on the relative amplitude of the two harmonic components of the average terminal velocity}'' from the perspective of the general formalism developed in~\cite{cuesta:2013,casado:2015}. In fact, what we do claim is that such a formalism in its lowest order approximation cannot explain (``is unable to account for'') an important aspect of such a dependence, specifically, ``the dependence of the location of the maximum velocity on $\omega \tau$''. Therefore, M\&C base their criticism on a misunderstanding or misrepresentation of what is actually claimed in~\cite{casado:2018}. Furthermore, the conclusion of the Comment~\cite{martinez:2021} is ambiguous and may mislead the reader into thinking that the theoretical explanations that are actually discussed in~\cite{casado:2018} (by using the adiabatic approximation and/or symmetry arguments) are incorrect. 

The aforesaid text of~\cite{casado:2018} is used by M\&C to criticize the general formalism developed in~\cite{cuesta:2013,casado:2015}. They claim that such a formalism is in general incorrect because it cannot explain the dependence of the optimal value of $\zeta$ that maximizes $\overline{V}_2(\zeta)$ on a new parameter $\alpha$ not present in the commented paper [compare Eq.~(2) of~\cite{martinez:2021} with Eq.~(13) of~\cite{casado:2018}]. Their argument is based on the wrong premise that ``\textit{the prediction coming from the aforementioned general formalism~[2,3]}'' (for Refs.~\cite{cuesta:2013,casado:2015}) is ``\textit{that the dependence of the average terminal velocity should scale as $\overline{V}_2(\zeta)\approx  C \zeta^2\alpha(1-\zeta)$}'' [see paragraph before Eq.~(5) of~\cite{martinez:2021}].  Actually, the general formalism developed in~\cite{cuesta:2013,casado:2015} predicts that the dependence of the average terminal velocity scales as $\overline{V}_2(\zeta)\approx  C \zeta^2\alpha(1-\zeta)$ only in its lowest order approximation, which is expected to be accurate only for small values of $f_0$. For the values of $f_0$ considered in~\cite{martinez:2021} ($f_0= 100$ and $f_0=1$), it is to be expected that the lowest order approximation is not sufficient and higher order terms are required. In fact, it is not necessary to introduce the new parameter $\alpha$, as M\&C do,  to realize that the lowest order approximation fails when the expansion parameter $f_0$ is not small enough. This issue is already mentioned in~\cite{casado:2018} when we say that the lowest order approximation ``is unable to account for the dependence of the location of the maximum velocity on $\omega \tau$'' observed in Fig.~3 of~\cite{casado:2018}. Therefore, the argument used by M\&C against the formalism developed in~\cite{cuesta:2013,casado:2015} is invalid because it is based on their confusion between the general formalism per se and its lowest order approximation. 

As a matter of fact, the arguments presented in~\cite{martinez:2021} demonstrate unequivocally that the supposedly universal results predicted by the theory of ratchet universality developed in~\cite{chacon:2007b,chacon:2010} are in general incorrect. Indeed, on the first page of~\cite{martinez:2021}, M\&C claim that ``\textit{For any $\alpha>0$, we shall argue that the maximum velocity is reached for $\zeta=2\alpha/(1+2\alpha)$ as predicted by the theory of ratchet universality (RU) [4-6]}.'' [see also Eq.~(4) of~\cite{martinez:2021}]. Therefore, for $\alpha=1$, which is the only value considered in~\cite{casado:2018}, the theory of ratchet universality predicts that the maximum velocity is reached for $\zeta=2/3$, irrespective of the particular value of the dimensionless relaxation time $\omega \tau$. Remarkably, in this case the theory of ratchet universality predicts the same optimal value of $\zeta$ as the lowest order approximation of the general formalism developed in~\cite{cuesta:2013,casado:2015}. This prediction is clearly incorrect because, according to the results shown in Fig.~3 of~\cite{casado:2018}, the optimal value of $\zeta$ that maximizes the velocity depends on the value of $\omega \tau$.

In the Comment~\cite{martinez:2021}, M\&C also propose an explanation of the supposed effectiveness of the ratchet universality prediction given by Eq.~(4) of~\cite{martinez:2021}. This explanation is untenable since it is based on an equation [specifically, Eq.~(7) of~\cite{martinez:2021}] that is clearly incorrect. As can be easily shown, Eq.~(7) of~\cite{martinez:2021} cannot be correct because it predicts an inconsistent result for the terminal velocity. Indeed, integrating Eq.~(7) of~\cite{martinez:2021}, one obtains that
\begin{equation}
	v_2(\theta)=-\frac{A a_0 \theta}{\omega \tau}+G(\theta),\label{error1}
\end{equation}
where, according to M\&C, $-A a_0$ is given by Eq.~(8) of~\cite{martinez:2021} and $G(\theta)$ is a $\pi$-periodic function of the dimensionless time $\theta= \omega t$ that can be expressed as a Fourier series. The terminal velocity can be obtained from Eq.~(\ref{error1}) by taking the long time limit $\theta \to \infty$. Thus, unless $A a_0\equiv 0$, which is not the case as can be seen in Fig.~4 of~\cite{martinez:2021}, Eq.~(7) of~\cite{martinez:2021} predicts an infinite terminal velocity. This prediction clearly contradicts the numerical results presented in~\cite{casado:2018} and~\cite{martinez:2021} and, consequently, Eq.~(7) of~\cite{martinez:2021} is incorrect. The statement made by M\&C that Eq.~(1) of~\cite{casado:2018} can be written as Eq.~(6) and (7) of~\cite{martinez:2021} is therefore also incorrect.

In addition, M\&C propose a possible explanation of the dependence of the location of the maximum velocity on $\omega \tau$ using the vibrational mechanics approach developed in Ref.~\cite{blekhman:2000}. In order for this approach to be valid, the frequency of the ``\textit{fast}'' force should be much higher than the frequency of the ``\textit{slow}'' force. However, in the case considered by M\&C, the frequency of the ``\textit{fast}'' force is only twice the frequency of the ``\textit{slow}'' force [see paragraph above Eq.~(12) of~\cite{martinez:2021}] and, therefore, the vibrational mechanics approach is not justified. Furthermore, although M\&C do not give an explicit derivation of Eq.~(12) of~\cite{martinez:2021}, it is easy to see that it cannot be correct. Indeed, because of the term $\smash{V_2^{3/2}}$, Eq.~(12) of~\cite{martinez:2021} leads to unphysical complex values of the velocity $V_2$ if the initial velocity is negative, which does not make sense. Therefore, the conclusions that M\&C draw from Eq.~(12) of~\cite{martinez:2021} are, to say the least, unreliable.  

It is worth mentioning that the asymptotic behavior $V_2\sim e^{-4 t/\tau}$ as $t\to \infty$ reported in~\cite{martinez:2021} is also incorrect.
It is easy to verify that Eq.~(12)  of~\cite{martinez:2021} can be written as
\begin{equation}
	\label{linear}
	-2 \frac{d }{dt_{\tau}}\left(V_2^{-1/2}\right)+V_2^{-1/2}=\frac{4 \pi}{\delta_0},
\end{equation}
which is a linear differential equation for $\smash{V_2^{-1/2}}$. The general solution of Eq.~(\ref{linear}) is $\smash{V_2^{-1/2}=4\pi/\delta_0+ C e^{t_{\tau}/2}}$, where $C$ is a constant that depends on the initial conditions. Taking into account that $t_{\tau}=t/\tau$ (see~\cite{martinez:2021}), one obtains that $V_2=(4\pi/\delta_0+C e^{t/(2 \tau)})^{-2}$. Therefore, the correct asymptotic behavior as $t\to \infty$ is $V_2\sim e^{-t/\tau}$ and not $V_2\sim e^{-4 t/\tau}$, as M\&C claim.

A second criticism raised by M\&C refers to the following sentence that appears in the second paragraph of the Introduction section of~\cite{casado:2018}: ``In this class of models, the mechanism behind the generation of directed motion is basically harmonic mixing~[3,8,12].'' (for Refs.~\cite{hanggi:2009,salerno:2002,marchesoni:1986}). In reference to this sentence, M\&C claim ``\textit{Finally, the author claims that: ``In this class of models [for rocking ratchets in the presence of thermal noise], the mechanism behind the generation of directed motion
	is basically harmonic mixing\dots .'' This is incorrect. \dots}''. Surprisingly, M\&C decide to leave out an important part of the sentence: the references~[3,8,12]. By hiding this information, they turn what is only a reference to previous results by other authors into a claim made by the author of~\cite{casado:2018}.  Again, their criticism is not aimed at a new result reported in~\cite{casado:2018} but at a matter, the rectification via harmonic mixing, that is not even used in~\cite{casado:2018} and that has been widely discussed in the literature (see, e.g., ~\cite{hanggi:2009,salerno:2002,marchesoni:1986}). Therefore, a Comment on~\cite{casado:2018} is not the right place to question this matter.

In conclusion, the Comment~\cite{martinez:2021} do not question any of the new results reported in~\cite{casado:2018}. On the contrary, the numerical simulations contained in~\cite{martinez:2021} corroborate the validity of the analytical expression for the average terminal velocity in the adiabatic limit derived in the commented paper [Eq.~(12) of~\cite{casado:2018}]. The authors of the Comment focus their criticism on two issues of little relevance to the commented paper: the general formalism developed in~\cite{cuesta:2013,casado:2015} and the rectification via harmonic mixing discussed, for example, in~\cite{hanggi:2009,salerno:2002,marchesoni:1986}. These two issues barely take up five sentences in~\cite{casado:2018}. Their criticism is based on a misunderstanding or misrepresentation of what is actually claimed in~\cite{casado:2018}. In particular, the argument presented in~\cite{martinez:2021} against the general formalism developed in~\cite{cuesta:2013,casado:2015} is invalid because it is based on a confusion by the authors of the Comment between the general formalism per se and its lowest order approximation. In fact, the arguments contained in~\cite{martinez:2021} do not show that the formalism developed in~\cite{cuesta:2013,casado:2015} is in general incorrect, as the authors of the Comment suggest, but rather that the supposedly universal predictions of the theory of ratchet universality developed in~\cite{chacon:2007b,chacon:2010} are in general incorrect. In addition, it has been shown that other arguments presented in~\cite{martinez:2021} are invalid because they are based on incorrect equations.

\begin{acknowledgments}
J.C.-P. acknowledges financial support from the Ministerio de Econom\'{\i}a y Competitividad of Spain through Project No. FIS2017-86478-P and from the Junta de Andaluc\'{\i}a.
\end{acknowledgments}

\end{document}